\documentclass[structabstract]{aa}  
\usepackage{graphicx}
\usepackage{natbib}
\bibpunct{(}{)}{;}{a}{}{,} 

\usepackage{txfonts}
\begin{document}
   \title{A method for quantifying the Gamma Ray Burst ``bias''. \\ Application in the redshift range 0-1.1}

 \author{S. Boissier, 
          \inst{1}
          \and
	R. Salvaterra
         \inst{2}
          \and
	 E. Le Floc'h
         \inst{3}
          \and
          S. Basa
          \inst{1}
          \and
          V. Buat
         \inst{1}
          \and
	 N. Prantzos
         \inst{4}
          \and
	S.D. Vergani
         \inst{5}
         \and
         S. Savaglio
         \inst{6}
          }
 \institute{Aix Marseille Universit\'e, CNRS, LAM (Laboratoire d'Astrophysique de Marseille) UMR 7326, 13388, Marseille, France 
	\and
INAF, IASF Milano, via E. Bassini 15, I-20133 Milano, Italy
	\and
Laboratoire AIM, CEA/DSM/IRFU, CNRS, Universit\'e Paris-Diderot, 91190 Gif, France
	\and
Institut d'Astrophysique de Paris, UMR 7095 CNRS, Univ. P. \& M. Curie, 98bis Bd. Arago, 75104, Paris, France
\and
GEPI, Observatoire de Paris, CNRS, Univ. Paris Diderot, 5 place Jules Janssen, F-92190 Meudon, France
\and
Max Planck Institute for extraterrestrial Physics, PO Box 1312, Giessenbachstr., D-85741 Garching, Germany
}

   \date{accepted in A\&A}

 
  \abstract
   {Long Gamma Ray Bursts (LGRBs) are related to the final stages of evolution 
of very massive stars. As such, they should follow the star formation 
rate (SFR) of galaxies. We can use them to probe for star-forming
galaxies in the distant universe following this assumption.
The relation between the rate of LGRBs in a given galaxy and 
its SFR (that we call the LGRB ``bias'') 
may however be complex, as we have good indications that 
the LGRB hosts are not perfect analogues to the general population of 
star-forming galaxies. }
   {In this work, we try to quantify the dependence of the LGRB bias on physical parameters of their host galaxy such as the SFR or the stellar mass. These trends may reveal more fundamental properties of LGRBs and their progenitors such as the role of the metallicity.}
   {We propose an empirical method based on the comparison of stellar mass functions 
(and SFR distributions) of LGRB hosts and of star-forming galaxies
in order to find how the bias depends on the stellar mass or the SFR.}
   {By applying this method to a sample of LGRB hosts at redshifts lower 
than 1.1, where the properties 
of star-forming galaxies are fairly well established, and where the properties 
of LGRB host galaxies can be deduced from observations (limiting ourselves
to stellar masses larger than 10$^{9.25}$ M$_{\odot}$ and SFR larger than $\sim$ 1.8 M$_{\odot}$ yr$^{-1}$), 
we find that the LGRB bias  depends on both the stellar mass and SFR. 
We find that the bias decreases with the SFR, i.e. we see no preference for highly star-forming galaxies,
once taken into account the larger number of massive stars in galaxies with larger SFR. We do not find
any trend with the specific star formation rate (SSFR) but the dynamical range in SSFR in our study is narrow.
Although through an indirect method, we relate these trends  to a possible decrease of the LGRBs rate / SFR ratio  with 
the metallicity.}
{The method we propose suggests trends that may be useful to constrain models of LGRB progenitors, showing a clear decrease of the LGRB bias with the metallicity. This is promising for the future as the number of LGRB hosts studied will increase.}

   \keywords{ Gamma-ray burst: general --Galaxies: evolution --
 Galaxies: high-redshift  -- Galaxies: mass function  --   Galaxies: star formation    }

   \maketitle
%

\def\msun{M$_{\odot}$}
\def\msunyr{M$_{\odot}$ yr$^{-1}$}
\section{Introduction}

The relation between Long duration Gamma Ray Bursts (LGRBs) and the explosion
of (very) massive stars is now established \citep[e.g.][]{woosley06}. 
Since massive stars are short-lived, it is often concluded that LGRBs can be used to
trace star formation up to very high redshift \citep[e.g.][]{kistler09,robertson12}\footnote{Short duration Gamma Ray Bursts (usually lasting less than 2 seconds) are not associated 
with massive stars and star formation; they are not considered in this work.}.
Owing to the relation between LGRBs and massive stars, it is possible to write that the rate of LGRBs in a galaxy is a simple function of its Star Formation Rate (SFR):
\begin{equation}
\dot{N}_{GRB}= b \times SFR,
\label{EQdefbias}
\end{equation}
where $b$ is the LGRB ``bias''. By this, we do not assume a priori that LGRB hosts  are 
necessarily biased with respect to field galaxies, but we define and study the relation between 
the two rates (SFR and LGRB Rate).
In an ideal case (assuming a constant initial mass function), the fraction of massive stars giving rise to LGRBs would be universal, and $b$ a constant.
This assumption is often implicitly made when the luminosity functions of ``normal''
star-forming galaxies and LGRB host galaxies are compared \citep[e.g.][]{lefloch03,basa2012}.
However, if $b$ were really a universal constant, the LGRB host galaxies should be
similar to galaxies selected by their SFR, and thus to star-gorming galaxies (SFGs).
On the contrary, it was found that 
host galaxies at low redshifts have lower luminosities and bluer colours \citep{lefloch03,fruchter99}, 
lower stellar masses
(e.g. Castro Cer{\'o}n et al. 2010, see however Perley et al. 2003,  Kr{\"u}hler et al. 2011),
lower metallicities
\citep{modjaz08,levesque2010,han10}, 
higher [Ne III] fluxes indicating very massive star formation \citep{bloom98},
more irregular morphology \citep{fruchter06},
or larger specific star formation rates \citep{christensen04,castro06} than SFGs.

Other approaches indicate that the bias must evolve with redshift, and that the 
LGRB rate is enhanced at high redshift with respect to
the SFR \citep[e.g.][]{daigne2006,kistler09,virgili11}.
Such studies combine hypothetical variations of the bias $b$ with models of the
intrinsic evolution of galaxies from the highest redshifts to the present day, and with the sensitivity of the instruments used to detect LGRBs. Their
predictions are then compared to the observed redshift distribution of LGRBs \citep[e.g.][]{salvaterra12,robertson12,elliott12}
or to other quantities such as the luminosity distribution \citep{wolf07}, 
or the mass function of LGRB host galaxies \citep[e.g.][]{kocevski09}.
This dependence of the LGRB bias on the redshift is sometimes written 
under the form:
\begin{equation}
b=b_0 (1+z)^n
\end{equation}
with $n$ usually found between $n$=0.4 and 1.2 \citep[e.g.][]{robertson12,qin10,kistler09}. 
Since a star at the end of its life does not know its redshift,
this evolution should be ascribed to changes in 
physical properties of galaxies varying with cosmic time. 
The culprits could be the initial mass function (changing the relative number of 
massive stars that could evolve into a LGRB in proportion to the 
SFR integrated over the full initial mass function as for other SFR tracers), or the metallicity.
A possible effect of metallicity on the occurrence of a LGRB 
has some theoretical support \citep[e.g.][]{macfadyen99,woosley06b,georgy09,pod10}. 
A metallicity effect was considered
under several forms (including cut-offs) and used to interpret 
various sets of data \citep[e.g.][]{wolf07,modjaz08,graham12}. 

We adopt the point of view that host galaxies should in fact form a subset of SFGs, and that the 
differences between this subset and the whole population of SFGs  is due to a 
dependence of $b$ on physical properties (deciding which of these galaxies 
hosts LGRBs or not). These relations should also be responsible for the apparent redshift evolution of 
$b$ when integrated over the whole population of galaxies.
We propose a relatively direct method to search for such effects. 
We quantify the differences between LGRB host galaxies and SFGs (in terms of 
the distribution functions of their stellar masses and SFR) in order to measure 
how $b$ depends on these and other quantities (such as the metallicity).
Section \ref{sec:forma} proposes the formalism and methodology 
to achieve such a goal, applied to the redshift range 0 to 1.1 for this study. 
The sample we selected to test our method is presented in Sect. \ref{sec:data}, while
our results are described in Sect. \ref{sec:appli}.
In Sec. \ref{sec:discu}, we discuss the main limit of our approach (the apparent 
dichotomy between SFGs and LGRB hosts). Our conclusions are
summarized in Sect. \ref{sec:conclu}.


\section{Formalism}
\label{sec:forma}

\subsection{Dependence of the LGRB bias on the SFR}
\label{methodSFR}
The number of star-forming galaxies with a given SFR is given by the
SFR distribution $\phi_{SFG}$. This function
is nowadays measured in a number of deep surveys, the SFR being derived
for instance from the rest-frame UV or the far-infrared luminosity
\citep{lefloch05,bell07,magnelli09,rodighiero10} or 
both \citep{martin05,bothwell11}. 
In a logarithmic form, this distribution is defined by:
\begin{equation}
  dN_{SFG} = \phi_{SFG}(logSFR) \ d(logSFR).
  \label{EQsfrfunc}
\end{equation}
The number of LGRBs occuring in each galaxy with a given SFR is proportional to
$b \times SFR$ (equation  \ref{EQdefbias}).
Combining this with equation \ref{EQsfrfunc}, we obtain the following expression
for the number of LGRBs  occurring in galaxies with a given SFR:
\begin{equation}
  dN_{GRB} \propto \phi_{SFG}(logSFR)  \times b \times  SFR \ d(logSFR).
\end{equation}
The SFR function of galaxies hosting LGRBs, $\phi_{LGRB}$ can in principle be
measured 
and has  the same form as equation \ref{EQsfrfunc} \emph{by definition}:
\begin{equation}
  dN_{GRB} = \phi_{GRB}(logSFR) \ d(logSFR).
  \label{EQLGRBhsfrfunc}
\end{equation}
As a result, if the SFR functions of SFGs and of host galaxies are
measured, then the dependence of the bias $b$ on the SFR can 
be directly found by combining equations \ref{EQsfrfunc} and \ref{EQLGRBhsfrfunc}:
\begin{equation}
b \propto \frac{1}{SFR} \times  \frac{\phi_{GRB}(logSFR)}{\phi_{SFG}(logSFR)}.
\label{EQbfromsfr}
\end{equation}
Note that 
we are only interested in relative trends with the SFR 
and not with the absolute value of the bias. We will then forget the 
normalisations of the functions in the following. 

To use equation \ref{EQbfromsfr}, a SFR distribution function for SFGs
is needed.  Although some differences are observed between various studies (due
to different SFR tracers or methodology used), the evolution
of this function is well documented in the redshift range 0 to 1.1
\citep[for a compilation of works, see][]{boissier10}.  For
simplicity, we will use the SFR distribution from the models presented
in \citet{boissier10}.
These simple models globally match the evolution of star-forming galaxies at 
redshifts lower than 1.1 \citep{buat08,boissier10} even if they slightly under-predict the number of very
active galaxies (SFR larger than about 30 \msunyr at high redshifts).  
\label{secfoncSFGs}
The advantage of
using models is that their evolution is smooth, free of small
variations that may be found in the observations performed at various
redshifts by different teams due to cosmic variance and
methodology.  Using such simple models eliminates this source of
``noise'' while keeping the general trends. It is also easy to
interpolate among the models in order to compute the SFR distribution
(and other quantities that will be useful in the remaining of this paper) for
SFGs at any redshift we are interested in.


\subsection{Dependence of the LGRB bias on the stellar mass}
\label{methodMSTAR}

The same exercise can be done by considering the stellar mass function of galaxies instead of the  SFR distributions. 
It is trivial to find that in this case, using $\tilde{\phi}$ for
the stellar mass functions (in logarithmic form):
\begin{equation}
b \propto  \frac{1}{SFR} \times \frac{\tilde{\phi}_{GRB}(log M{*}/M_{\odot})}{\tilde{\phi}_{SFG}(log M{*}/M_{\odot})}.
\label{EQbfromMstar}
\end{equation}
Here, $b$ depends on the ratios of the stellar mass functions for a given stellar mass, 
but also on the SFR (since we are interested only on the trends, the normalisations of the functions 
will be neglected). 
However, by restricting ourselves to a
redshift range where a stellar mass-SFR trend exists (see next section), we 
will be able to derive the SFR from the stellar mass, and thus obtain a dependence 
of $b$ on the stellar mass uniquely.

The stellar mass function of SFGs needed in the computation  is well constrained
in the redshift range 0 to about 1 down to  $\sim$ 10$^9$ \msun{} 
\citep{ilbert10}. Moreover, the stellar mass function of SFGs evolves very little in this
redshift interval \citep{borch06,arnouts07,bell07,cowie08,vergani08,pozetti10,ilbert10,ilbert13}.
For simplicity sake, we will use again the model stellar mass function from 
 \citet{boissier10}, with the advantage of being rigorously constant in the redshift range 0 to 1.1.

   \begin{figure}
   \centering
   \includegraphics[width=7cm]{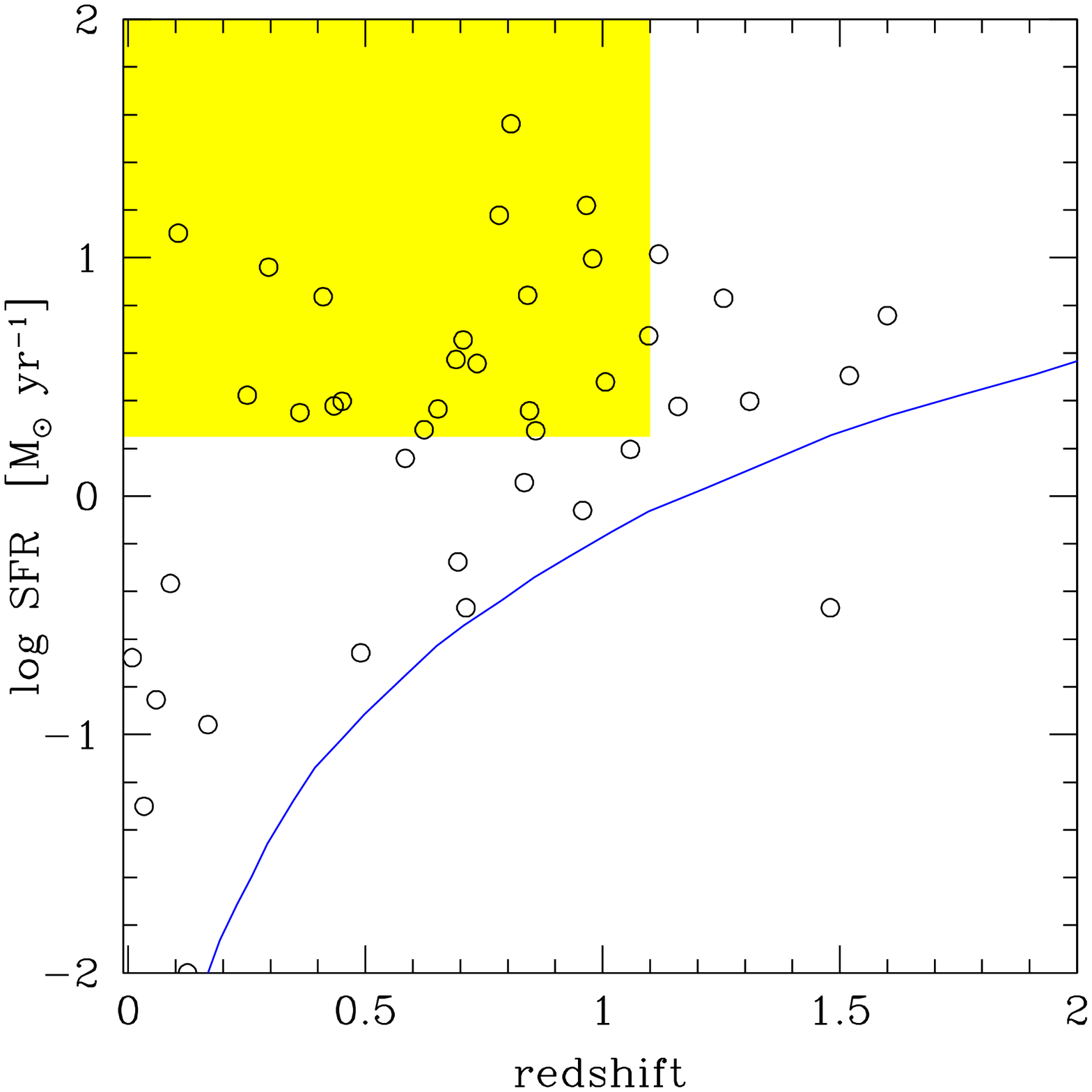}\\
   \includegraphics[width=7cm]{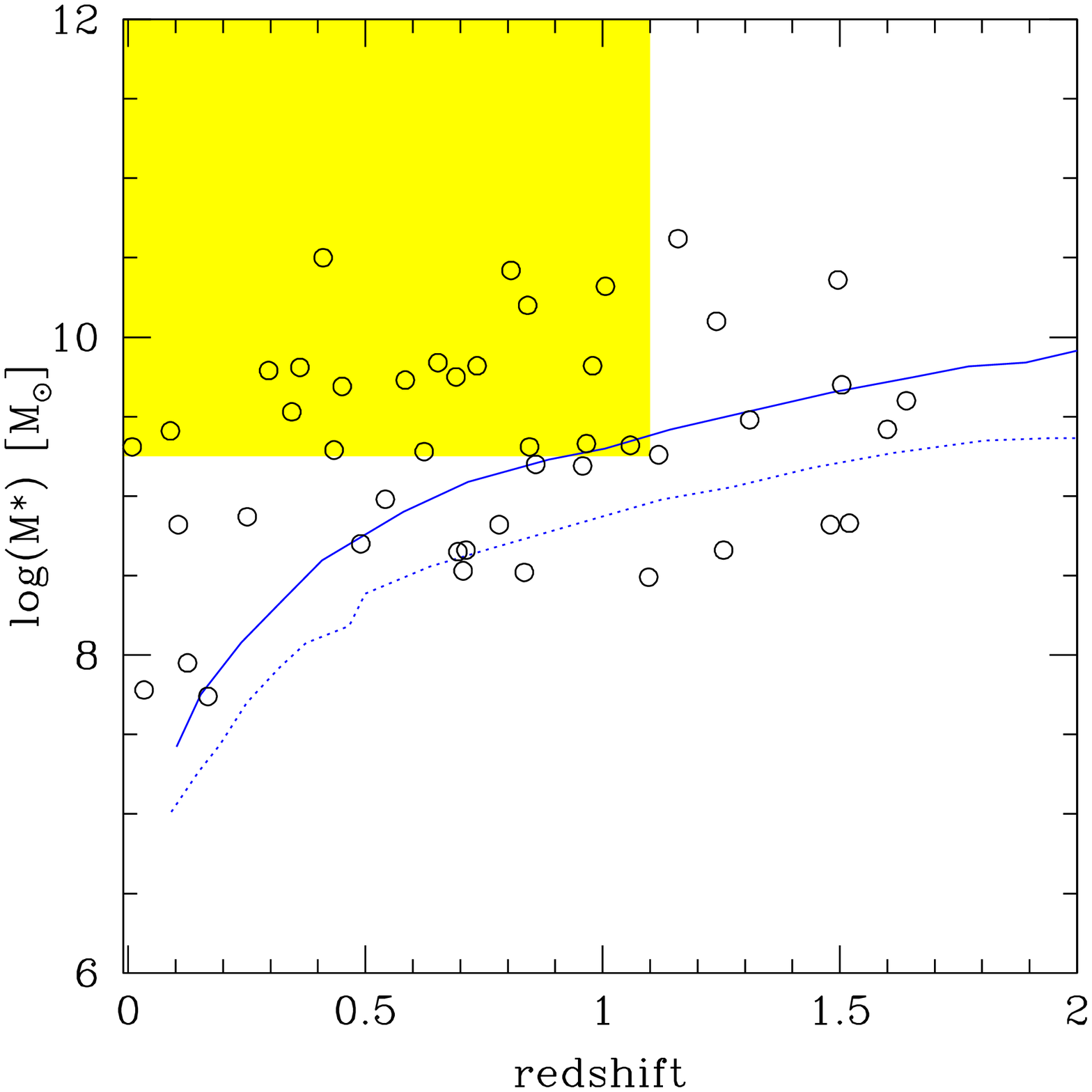}
   \caption{Galaxy stellar mass (bottom) and SFR (top) for LGRB host galaxies considered in this paper. The lines indicate the limits shown in the figures of \citet{savaglio09}:
the solid and dashed lines in the bottom panel show the stellar mass
as a function of redshift of a galaxy with a K-band magnitude of 24.3, and old stellar population
or constant SFR, respectively.
In the top panel, the line represents an H$\alpha$ or [O II] emission flux of 1.3 $\times$ 10$^{-17}$ erg s$^{-1}$ or 0.7 $\times$ 10$^{-17}$ erg s$^{-1}$,
respectively, assuming a dust extinction in the visual band A(V) = 0.53. 
The shaded area indicates our selection criteria: 
we work at redshift lower than 1.1 and above a minimal value for the stellar masses and SFR so that our sample is complete. }
    \label{FigSelec}%
    \end{figure}

\subsection{Relationships among galactic properties}

The methods described in sections \ref{methodSFR} and
\ref{methodMSTAR} provide constraints on the variation of the
LGRB bias $b$ on the SFR or on the galaxies stellar mass.  However, such a
dependence does not
necessary imply a physical relationship. Indeed, these quantities may
themselves be correlated to more fundamental ones for the physics of
LGRBs (such as the metallicity). Important relations that we will use
to analyse our results are :
\begin{itemize}
\item the stellar mass - SFR relationship.  The existence of a stellar
  mass - SFR relationship at all redshifts may be debated, but several studies
  indicate a good relation between the two quantities at redshifts lower than about 1
 at least from a statistical point of view despite some dispersion
  \citep{brinchmann04,buat08,gilbank2011,salmi12,berta13}. 
This relation is necessary  to use the method described in
 section \ref{methodMSTAR}.  Moreover, in combination with our results ($b$ -
 stellar mass, and $b$ - SFR relationships), it can be used to attempt to
 determine trends with the SSFR. In practice,
we will use the stellar mass - SFR relationship of the models of \citet{boissier10}, in broad agreement
with the observed trends in the redshift range 0 - 1.1. We refer the reader to sect.
\ref{sec:discu} for a discussion including the role of the dispersion in this relation.

\item The stellar mass - SFR - metallicity relationship.
\citet{mannucci10} found a ``Fundamental Metallicity Relation'' (FMR)
between the metallicity and a combination of the stellar
mass and the SFR \citep[see also][for a similar
 relationship]{laralopez2010}. This relation presents a smaller
scatter than traditional mass-metallicity relationships 
 and has the advantage of being independent of redshift (at least
at redshift lower than 2.5). Moreover, the relation also holds
for GRB hosts \citep{mannucci11}.
Once a trend between $b$ and the SFR (or $b$ and the stellar mass) is established,
we can use the stellar-mass SFR relationship to compute the stellar mass from the SFR (or vice-versa) and thus find a trend between $b$ and the metallicity simply assuming 
the \citet{mannucci10} relation. 
\end{itemize}

   \begin{figure*}
   \centering
 \includegraphics[width=14cm]{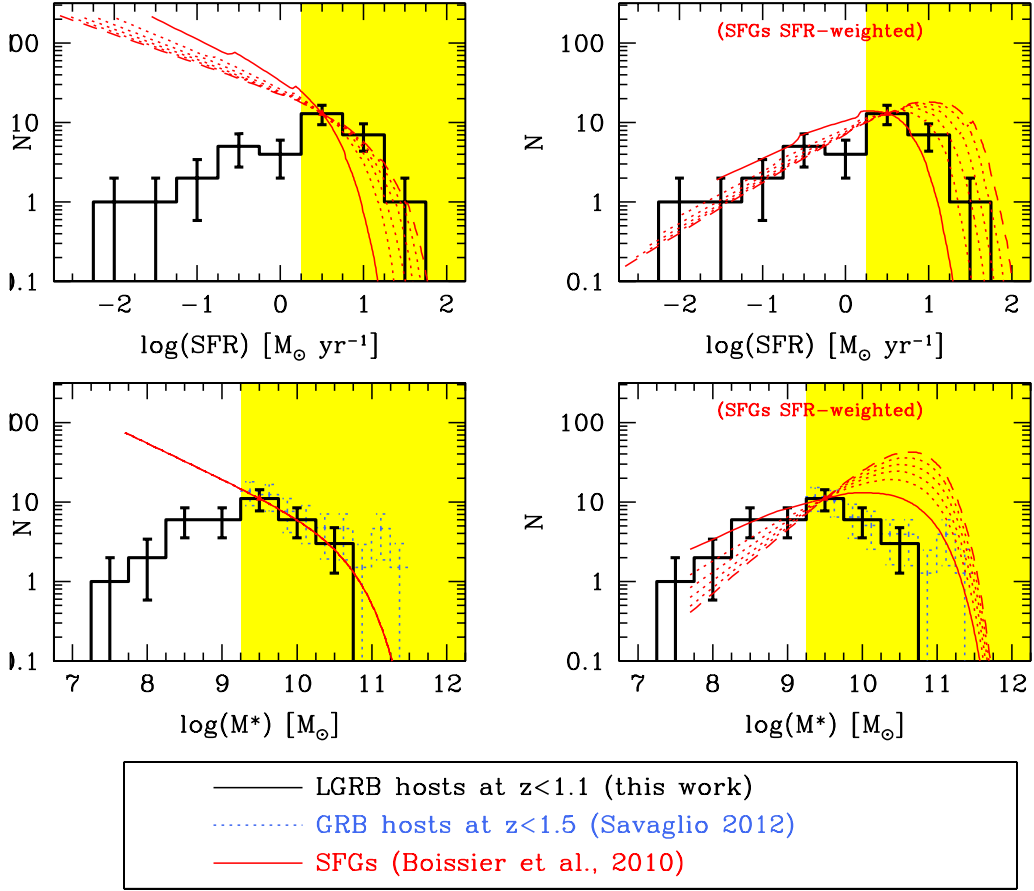} 
   \caption{
The histograms show the stellar mass function (bottom) and SFR distribution (top) of the LGRB host galaxies. 
The shaded area indicates the region where we expect to be complete. Outside of this region, our results are likely to underestimate the number of host galaxies.
In the bottom panels, the dotted histogram reports the stellar mass function of LGRB hosts by Savaglio (2012) as a comparison. In left panels, the LGRB host distributions are compared with those of star-forming galaxies as modelled in Boissier et al. (2010). Solid and dashed curve refers to redshift 0 and 1.1, respectively, while dotted lines are for intermediate redshifts (0.3, 0.5, 0.7, 0.9). As the LGRB rate is proportional to the SFR (for constant $b$) a direct comparison between LGRB host and SFG distributions is not possible. Right panels show the same comparison when SFG distributions are weighted by the SFR. In this case, for a constant bias $b$, the LGRB host and SFG distribution should be identical.
}
    \label{FigHOSTfunctions}%
    \end{figure*}

\section{LGRB hosts Sample}

\label{sec:data}

\subsection{Selection}

\label{sampleselection}

To apply the methods described in section \ref{sec:forma}, we need to
know the distribution of stellar masses and SFRs of LGRB host
galaxies.  For this first application of the method, we try to
determine such functions from the data available in the GHostS
database (as on the 9th of July 2012).  Fig. \ref{FigSelec} shows the
SFR and stellar masses (as taken from the GHostS database) of the
LGRB hosts as a function of redshift.
The separation between short and long  LGRBs is not uniquely defined \citep[e.g.][]{zhang12}.
For the present work, we removed all the bursts considered as short by \citet{kopac12}, and the
bursts with duration shorter than 2 seconds among the remaining ones.
We limit our study to redshift lower than 1.1 where the \citet{boissier10} models 
used in our analysis represent a good description of the SFG population.
Since the database is a compilation of all hosts known with public
information, there is unfortunately not a clear limit on the stellar mass or SFR to
adopt. 
However, the stellar masses derived from the hosts SED in
\citet{savaglio09}, providing a large number of the data in the GHostS
database, correlates well with the K band magnitude. The K band
limit shown in Fig.  \ref{FigSelec} provides a good idea of the
minimal stellar mass that can be measured as a function of redshift. 
Similarly SFR are derived from a variety of SFR
tracers, but the curve in Fig.  \ref{FigSelec} taken again from
\citet{savaglio09} provides a good idea of the measurement limit in
SFR. The SFRs from \citet{savaglio09} are corrected for dust attenuation
using adequate extinction tracers when they are available or the average 
value of 0.5 magnitude when they are not.
Based on our redshift range ($z<1.1$) and these limits, 
we obtain an usable dynamical range by including
galaxies with $log(M{*}/M_{\odot})>9.25$ for the study based
on the mass \citep[the same limit adopted in][]{savaglio12}. 
Independently we construct a sample of galaxies with
$log(SFR/1 M_{\odot}yr^{-1})>0.25$ (i.e. SFR larger than about 1.8 \msunyr) for 
the study based on the SFR. This limit is quite conservative to
compensate for the large uncertainties in SFR measurements.   
Of the 66 galaxies with measured stellar masses in GHostS,
35 LGRB hosts are found with $z<$1.1 (20 above our stellar mass limit).
Of the 48 galaxies with measured SFR in GHostS, 
34 LGRB hosts are found with $z<$1.1 (21 above our SFR limit).

\subsection{Stellar mass and SFR distributions}
\label{LGRBdistribs}

The histograms of the measured stellar masses and SFR in LGRB hosts,
within our selection limits are  shown
in Fig. \ref{FigHOSTfunctions} (as a sanity check, we randomly split our
sample in two halves and obtained consistent distributions).
In this figure, a turnover is observed below our selection limit, what
shows indeed that we are missing
LGRB hosts below this limit. In the rest of the paper we will then keep only the 3 bins for the
largest values of the SFR and the stellar masses, 
for which we believe to be complete.
The distribution of stellar masses and SFR in 
SFGs \citep[from the models of][]{boissier10} are also shown.
A direct comparison between hosts and SFG distribution is not possible. Indeed, even in absence of bias (constant $b$), hosts should present larger SFR than SFGs, since the LGRB rate is proportional to the SFR. As shown by equation  \ref{EQbfromsfr}, the bias $b$ is proportional to the ratio of the two functions divided by the SFR.
In order to obtain a meaningful comparison, we also show in the top-right panel of  Fig.  \ref{FigHOSTfunctions} the SFR distribution of SFGs weighted by their SFR. In the absence of bias, the SFR weighted distribution for 
SFGs and the LGRB hosts distribution should be identical. A similar effect applies to the stellar mass functions. We show in the bottom-right panel the stellar mass function of SFGs weighted by their SFR.
In the bottom panel, we compare our stellar mass distribution with the stellar mass function
of \citet{savaglio12} for GRB hosts at redshift lower than 1.5. 
Despite the slightly different
selections, the mass functions are consistent with each other within the statistical uncertainties.
The main difference is found for the largest stellar masses. As a test, we added in our analysis 
a bin centred on 10$^{11}$ \msun{} including 2 fake hosts. With this bin, our distribution
would be in perfect agreement with  the \citet{savaglio12} mass function. Except for this new
bin, the results presented in the rest of the paper would be of course unchanged. 
The only difference would then be the addition of one extra point corresponding to  10$^{11}$ \msun{} 
for which we would obtain values of $b$ similar to the one at 10$^{10.5}$ \msun{} with large error-bars (due to 
the small number statistics), and still in agreement with the overall trends found in the paper.
In other words, the difference between the \citet{savaglio12} distribution and ours could easily be
explained by the absence of 2 bursts due to Poison noise. Adding them artificially leaves our conclusions unchanged.

The binning was chosen in order to have at
least 10 LGRBs in the bin with the largest number of objects and still
a few objects in higher stellar masses and SFR bins.
Above our completeness limits, we can consider the histograms as the
true LGRB hosts mass function and SFR distribution (except for the
normalisation since we are interested only in relative trends).  This is very different from the usual way to
determine such functions from galaxy surveys for which galaxies of a
given luminosity are detected only in a limited volume, and volume
corrections have to be applied. In our case, the inclusion of the
galaxy in our sample is not dependent on its luminosity or on the
volume probed (as soon as it is possible to measure the SFR or the stellar mass).
Indeed the selection is made by the fact that a LGRB is
detected and correctly localised  what does not depend on its distance or on its host
luminosity (excluding only ``dark bursts'' which are discussed in the next section). 
Then the hosts are looked for and studied whatever is the
redshift.  We avoid volume corrections by selecting a
part of the space parameter (redshift, stellar mass, SFR) where the
SFR and stellar masses are sufficiently large to be measured up to the
largest redshift used.

\subsection{Dark bursts}

The functions obtained in the previous section could suffer from an
observational bias. They are constructed only for host galaxies that
were observed.  The hosts of ``dark bursts'' with no optical
afterglows, representing at maximum 30 \% of the LGRBs
\citep[e.g.][]{melandri12} are usually not identified and may be
absent from our sample. There are actually several definitions for
``dark bursts'' \citep[e.g.][]{greiner11}, and many LGRB hosts are
found from their X-ray afterglow even without optical afterglow \citep[e.g.][]{hjorth12}
so that the fraction missing from our sample is hard to determine but probably lower than
the number quoted above.
\citet{melandri12} find a similar redshift distribution between dark bursts and
the general population of LGRBs, suggesting our low redshift 
sample is also affected by dark bursts (even if their redshift distribution at redshifts 
lower than 1 is constrained by only very few objects).

It has been suggested that most of dark 
bursts are due to high amount of dust extinction in the proximity of the LGRB
making it too faint to be observed \citep[][]{melandri12,rossi12}. 
If dark bursts are indeed due to dust attenuation, it is expected that
their relative number would vary with the SFR and stellar mass of their host galaxies.
Indeed, \citet{perley13} and \citet{kruhler11} found that the host galaxies 
of very dust-attenuated LGRBs show higher SFR and are more
massive (again, few of their bursts are at redshifts lower than 1, thus
it is not completely obvious if this applies for the nearest hosts).
On the other hand, \citet{michalowski12} found recently that an optically 
unbiased sample of host galaxies at redshift lower than 1 has globally similar
SFR and amount of attenuation as normal galaxies.

As a summary, even if the situation is not completely clear, an increasing fraction of 
LGRB hosts with larger stellar mass and SFR may be missing from our sample. Our results
could then not result exclusively from  a real -physical- dependence of $b$ on various parameters, but
rather quantify how the dark burst bias depend on physical quantities in usual samples of hosts.
The recent study of dark bursts hosts by \citet{perley13} still suggests that a physical dependence is needed.

%

\section{Application in the redshift range 0 - 1.1}

\label{sec:appli}

\subsection{The bias - SFR and bias - stellar mass relationships }

   \begin{figure}
   \centering
   \includegraphics[width=7cm]{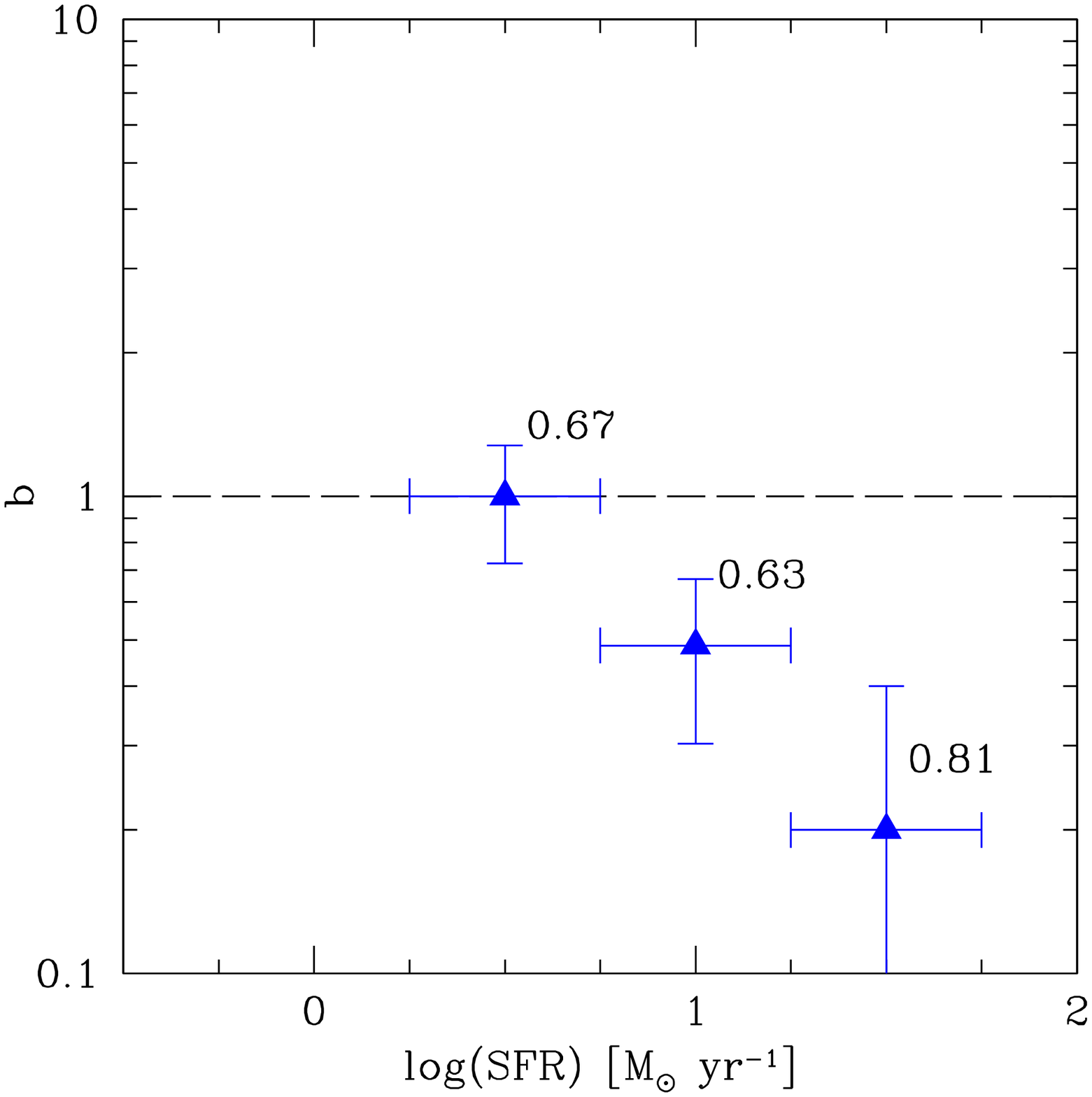} \\
   \includegraphics[width=7cm]{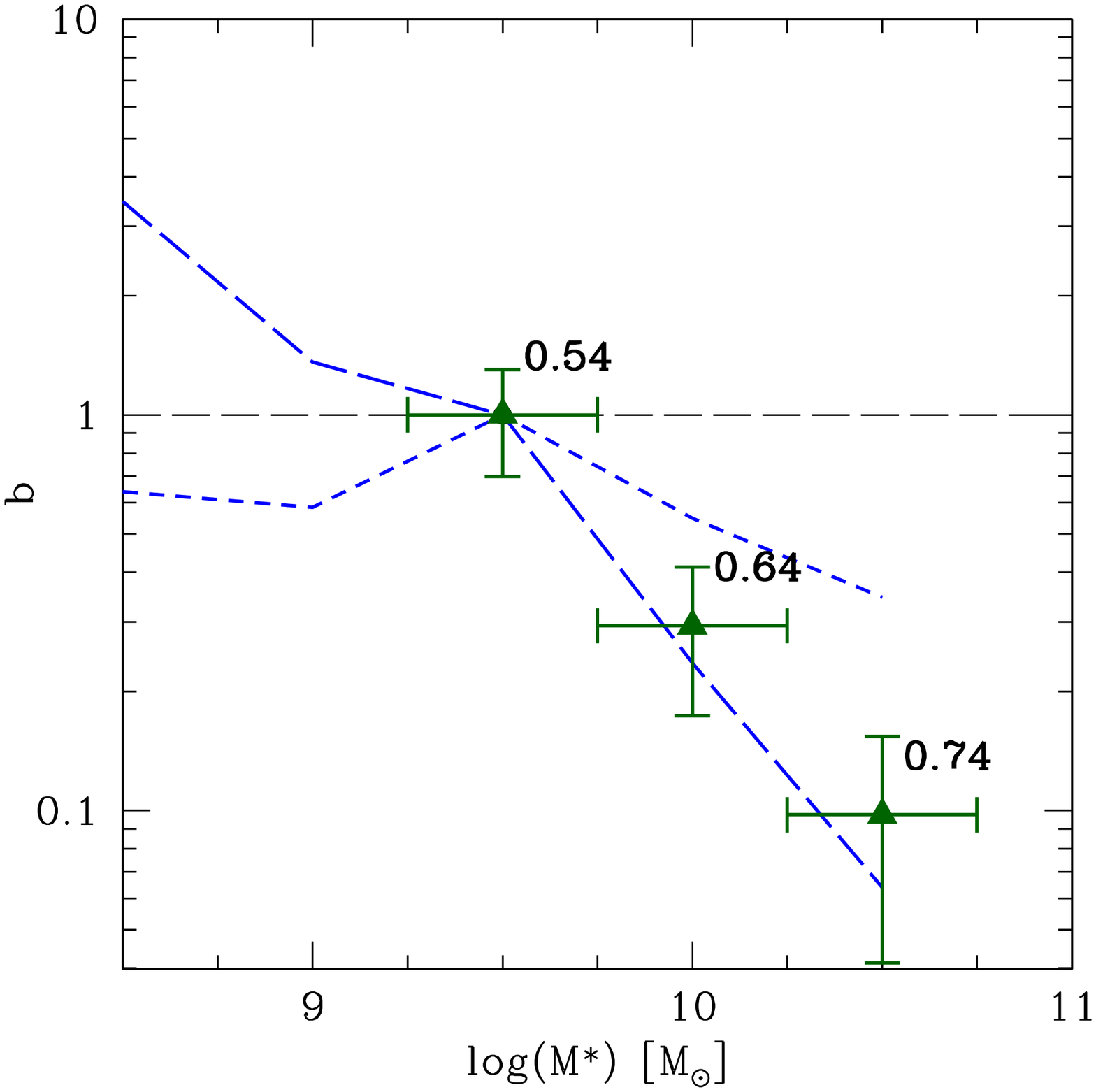}

   \caption{Variation of the bias $b$ with the stellar mass (bottom) and the SFR (top) with an arbitrary normalisation. 
The triangles show our results after interpolating the SFGs properties at the median redshift of the bin (indicated next to each symbols).
In the bottom panel, the dashed lines indicate the results adopting the two regression lines of the stellar mass-SFR relation found in host galaxies (shown in Fig. \ref{FigSFRmass}) as explained in Sect. \ref{sec:discu}. 
The vertical error-bars indicates the statistical uncertainty on the number of LGRBs ($\sqrt N$ in each bin). The horizontal error-bar only indicates the width of the bin.}
    \label{FigBIASconstraints}%
    \end{figure}

With the stellar masses and SFR distributions of both SFGs and LGRB
hosts (shown in Fig. \ref{FigHOSTfunctions}) it is
easy to derive how the bias $b$ depends on the SFR and stellar mass. 
Following equations \ref{EQbfromsfr} and \ref{EQbfromMstar}, 
the first step is to divide the distribution corresponding to 
host galaxies  by the one corresponding to SFGs. 
For the stellar mass function, this can directly be done (since the
mass function is constant in the redshift range). The SFR distribution
of SFGs on the other hand depends on the redshift (as can be seen in
Fig. \ref{FigHOSTfunctions}).  It was thus interpolated to the median
redshift of the bin before performing the division in each SFR bin.

The next step consists in dividing by the SFR. 
For the method based on the stellar mass function, we have to adopt
a stellar mass - SFR relationship that will give us the SFR for each stellar mass
bin. This relation for SFGs evolves with redshift according to the models in 
\citet{boissier10}. Here again, we interpolate in each stellar mass bin
to find the corresponding SFR for the median redshift of the bin.

Fig. \ref{FigBIASconstraints} shows the obtained relation between $b$ and the stellar mass (bottom panel) and the SFR (top panel). 
While indication about the existence of a bias in the LGRB host population has been inferred from their properties (blue colors, low metallicities, etc.), our method allows us for the first time to quantify its dependence with SFR and stellar mass.

A similar trend as the one presented in Fig. \ref{FigBIASconstraints}  between the bias and the stellar mass would be found 
by adopting the models at any fixed redshift between 0 an 1 rather than
the median redshift in the bin. This indicates that the redshift 
distribution in each bin does not strongly influence our results. 
We tried different binning schemes for the 
stellar mass histogram (larger/narrower, and shifted by a factor 1.5), and a similar 
trend was always found.

The trend between the bias and the SFR is less robust, as
the redshift evolution of the SFGs SFR distribution introduce a strong
dispersion in our results. Especially, our results are sensitive to
the particular redshifts of the LGRBs in each bin. We should also note
that the point in the highest redshift bin is poorly constrained: there 
is only one LGRB in this SFR bin and the SFGs SFR distribution for this high 
SFR may be under-estimated  (see section \ref{secfoncSFGs}) so that $b$ is over-estimated.
As a result of these effects, the trend with the SFR slightly depends on the binning scheme.
In most cases, a decrease of $b$ with the SFR is still suggested  but the relation is 
weaker for large bins (1 dex).

Combining the trends found between $b$ and the stellar mass (or with the SFR) and
the stellar mass-SFR relationship, it is straightforward to derive how $b$
depends on the SSFR.  Several studies found that 
LGRB hosts tend to have larger SSFRs than field galaxies
 \citep[e.g.][]{christensen04,castro06}. 
Surprisingly, We do not find a clear trend of $b$ with the SSFR. 
However, the very small dynamical range of SSFR probed by our sample,
less than 0.5 dex, and the uncertainties in the determination of both stellar mass
and SFR prevent us to draw any conclusion about this issue.

   \begin{figure}
   \centering
   \includegraphics[width=7cm]{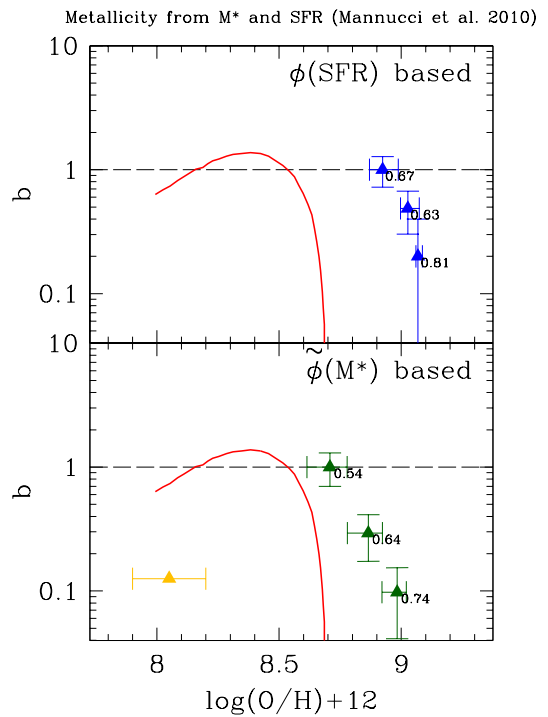}
   \caption{Relation between $b$ and the metallicity 
derived from the empirical FMR relationship of \citet{mannucci10}.
The vertical error-bars correspond to the statistical uncertainty on the number
of LGRBs. The horizontal error-bar attached to each symbol corresponds to the width of the bin
used at the beginning of the method. The horizontal bar in the bottom-left part of the figure
indicates a systematic uncertainty on abundance measurements of a typical
factor 2.
The top panel shows the results based on the SFR distribution function, and the bottom panels the results based on the stellar mass functions. 
Finally, the curve indicates the fraction of SN Ic-WO supernovae
(with which LGRBs could be associated)
according to  the models of \citet{georgy09}.
}
    \label{FigBIASmetal}%
     \label{FigBIASmetalMannucci}%
\end{figure}

\subsection{The bias as a function of the metallicity}

\label{secusingmannucci}

Combining the trends found between $b$ and the stellar mass (i.e. with the
method based on the stellar mass function), and the
stellar mass-SFR relation interpolated at the median redshift of each bin,
we can obtain a list of ($b$, stellar mass, SFR) triplets. An other set of 
such triplets can be obtained starting from the trend between $b$ and the 
SFR (i.e. the method based on the SFR distribution).
The  FMR of \citet{mannucci10} then allows us to compute 
from the SFR and stellar mass in each of these triplets
the metallicity of the galaxy.

The relations between $b$ and the metallicity obtained in this way 
are shown on Fig. \ref{FigBIASmetal}.
%
When we start from the $b$-stellar mass
relationship, a clear trend with metallicity is obtained.  It is also
suggested starting from the $b$-SFR relationship but the uncertainties
are larger and the dynamical range smaller.

These trends are compared to the predictions of \citet{georgy09} for 
SN Ic-WO (SN Ic with progenitors consisting of Wolf-Rayet stars with 
carbon surface abundance superior to nitrogen abundance, and
  C+O to He ratio in number larger than 1).  \citet{georgy09}
suggests a fraction of these SN could give rise to LGRBs.
They predict a rate that is
function of the metallicity and shown as the curve in Fig.
\ref{FigBIASmetal} (with an arbitrary normalisation).
In the case of their model, the metallicity is not measured but a physical parameter. 
Thus the comparison suffers from the large uncertainties in the calibration of metallicity indicators, \citep[e.g.][]{kewley08}. We
indicate in the figure a typical 0.15 dex uncertainty corresponding to a  factor 2 total variation for illustration purpose.
In relative terms, the drop with metallicity found by \citet{georgy09} is slightly stronger than the trend found
in our method based on the stellar mass (the method that is better constrained).
On the absolute scale, we find the decrease of $b$ at much higher metallicity than they do but we remind the reader that our determination
of the metallicity is based on the FMR relation that is quite dispersed. If LGRBs prefer low-metallicity, they will be found at the low-metallicity end of the scatter. Full models taking into account the scatter in SFGs could help us to test this possibility, but are beyond the scope of the simple approach proposed in this paper.

Our progressive decrease of the bias is at odds with the the idea
of a simple metallicity cut-off at a metallicity ten times lower than solar
proposed e.g. by \citet{niino09} on the basis of the Ly$\alpha$
emission of LGRB host galaxies statistics.  This trend agrees with
the analysis of \citet{campisi11} who found no need for a
low-metallicity threshold \citep[see also][]{levesque10b,graham12}.
These comparisons illustrate that our method already provides a
quantification of the bias-metallicity relationship that can be
compared to other recent theoretical or empirical works.
Nevertheless, our results will have to be confirmed with higher
statistics and better defined samples in the future.

\section{Discussion: on the enhanced SFR in LGRB hosts}

\label{sec:discu}

\begin{figure}
   \centering
   \includegraphics[width=7cm]{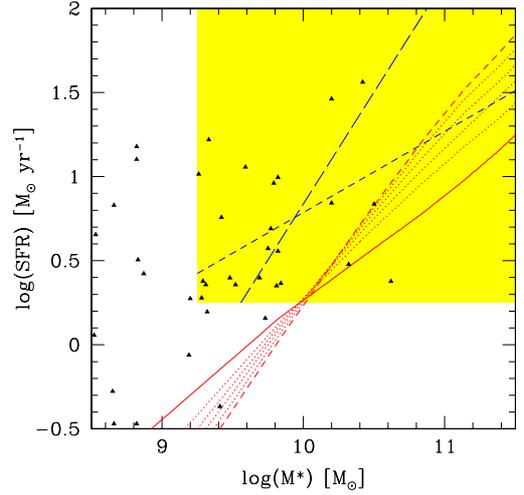}
   \caption{Stellar mass-SFR relation of LGRB hosts (symbols), compared to the
one taken from the models of \citet{boissier10} for SFGs shown as a set of curves 
(solid at redshift 0, dashed at redshift 1.1, dotted for intermediate redshifts). 
The shaded area corresponds to 
our completeness limits. The dashed lines within this area are the 
two regression lines (for the SFR as a function of the stellar mass, and the stellar mass as a function of the SFR) computed from the LGRB hosts data.}
    \label{FigSFRmass}%
    \end{figure}

The relation between the SFR  and the stellar mass observed in LGRBs differs from the one observed 
in SFGs, since LGRB host galaxies have higher SSFR \citep{christensen04,castro06}. 
While we argued that SFGs should be the parent population of galaxies
providing  host galaxies, we cannot reproduce the stellar mass-SFR relationship of 
LGRBs with the models of normal SFGs. This can be seen even in our sample in
Fig. \ref{FigSFRmass} where  the  \citet{boissier10} models are found below LGRB hosts. 
This figure may seem a bit different from the picture emerging from Fig. 12 of \citet{savaglio09}
in which the stellar mass-SFR relation of a sample of GRB hosts seems similar to the one observed in  samples of massive star forming galaxies and Lyman-break galaxies at high redshift. 
The differences between their result and our study are
i) the selection of high-redshift ``typical'' galaxies (e.g. Lyman-break galaxies) may lead to a sample not representing the
underlying star-forming population as a whole (bias towards active galaxies),
ii) our study is limited to redshift lower than 1.1, where less extreme objects in terms of SFR are usually found,
iii) the models represent typical star-forming galaxies, but may miss the more active ones (see Sect. \ref{secfoncSFGs}).

This last point is related to the fact that the
models have ``smooth'' star formation histories, without any dispersion
in the SFR at a given stellar mass. 
On the contrary, during their evolution, galaxies may suffer episodically 
increases of their SFR  (associated with e.g. interactions or mergers), that would bring them 
above the underlying stellar mass - SFR relationship (thus creating dispersion in
this relation). 
Because LGRBs are related to massive stars, LGRB host galaxies are more 
likely to be found among such starbursts (even  if $b$ is constant), which is a simple way
to explain the enhanced SSFR in LGRB host galaxies. 
In other words, as galaxies present a scatter of SSFR, LGRB hosts will be  
preferentially found at the upper end of the SSFR distribution, even if the SSFR by itself has
no influence on the occurrence of a burst.
A drawback of our method is that such a dispersion is  not taken into account in
the models of SFGs. 
However, if variations of the SFR history affect all galaxies 
independently of their stellar mass, the general effect would 
be simply a shift in the stellar-mass - SFR relationship
towards larger SFR during the events, 
what could produce the systematic shift between SFGs and host galaxies 
observed in Fig. \ref{FigSFRmass}.
We actually tested how much our results could be affected by 
the fact that we relied on the stellar mass-SFR relation of SFGs that
is lower than the one observed in host galaxies. 
To this aim, we fit the relation found in LGRBs 
(the two regression lines are shown in Fig. \ref{FigSFRmass}).
We then added in Fig. \ref{FigBIASconstraints} the $b$-stellar mass
relations obtained by adopting these fits of the stellar mass-SFR relation
in LGRB host galaxies (but without any information
concerning its redshift evolution) rather than the one from SFGs: the two
dashed lines in Fig.  \ref{FigBIASconstraints}  correspond to the two regression lines in Fig.
\ref{FigSFRmass}. Indeed, we obtain trends consistent with the one
found with the SFGs stellar mass-SFR relationship. 
The decrease of $b$ with the stellar mass above
10$^9$ \msun{} is a robust result despite this difficulty.

Recent simulations show that interactions and mergers may
indeed increase temporarily the star formation efficiency \citep[by up
  to a factor 10 according to][]{teyssier10}.  Simulations also
indicate that during interactions, low metallicity gas may be brought
towards the inner part of galaxies diluting its metallicity
\citep[e.g.][]{montuori10,rupke10,perez11}.  There are 
observational indications of reduced metallicities in merging pairs
\citep{kewley06,leo08} or galaxies with elevated merger-induced star
formation \citep{rupke08}.  \citet{montuori10} show examples of their
models in their Fig. 1 where a large increase of the gas in the inner
galaxy leads to a large SFR increase (by a factor up to 10), and a simultaneous 
dilution of the metallicity (from 0.1 to 0.3 dex). 
The metals produced in stars created during this peak of star formation
will enrich the gas and future generation of stars, 
but the stars themselves created during the event have
a diluted metallicity.
This simultaneous increase of the SFR and decrease of the metallicity
would make the galaxy stay at approximately the same spot in the
FMR relation of \citet{mannucci10}: for a 10$^{10}$ \msun{} galaxy with a
SFR going from 1 to 10 \msunyr, the metallicity given by the fit from
\citet{mannucci10} would decrease from 8.9 to 8.75 : a similar
change to the metal dilution found in the models.  While it would be
difficult to take into account in our approach all the possible interactions, these
increased SFR (favouring LGRBs in numbers), and dilution of the
metallicity would simply explain the tendency for host galaxies to be
shifted with respect to normal galaxies towards lower metallicities
and higher SFR \citep[see also ][]{kocevski11}.
If this happens for galaxies of all masses (or all SFR before the interaction) in a similar manner, 
the effect would  be systematic, and thus it would not erase  possible trends of $b$ with galactic
properties little affected by this event, such as the stellar mass.
In this case, our simple method should indeed provide correct results, 
even if such event are not (yet) included in the models we used.

In the future, we will include scatter in the models of the evolution
of star-forming galaxies (due to interactions, mergers, environment)
to see if the trends found with the simple method are still recovered,
together with the simultaneous enhancement of the SSFR and the
reduction of metallicity in host galaxies.

\section{Conclusion}

\label{sec:conclu}

The main objective of this paper was to present a simple ``empirical'' method
to measure how the bias $b$ (the ratio between the LGRB rate per 
galaxy and the SFR) may depend on physical quantities characterising the 
host galaxies (and thus the environment of the exploding massive star).

The method is based on the comparison of the stellar mass function and  
SFR distribution for LGRB host galaxies and for star-forming galaxies, 
the expected parent population of LGRBs.
For simplicity, we adopt the simple simple model of \citet{boissier10} for this family.
We work within a redshift range (and a parameter space) where all quantities are
well known. We also take advantage of the fact that star-forming galaxies
exhibit known relationships allowing to link easily various parameters (stellar masses,
SFR, and metallicity).
Given the statistics of LGRB events, a wide redshift range has to be considered, while some properties
of galaxies evolve during the corresponding time interval. Within the redshift range
where the relations are known, it is still possible to interpolate to the median
redshift of the data to partially eliminate this problem. 

When applied to a current sample of LGRBs at redshifts lower than 1.1,
with host galaxies of stellar masses larger than 10$^{9.25}$ \msun{} and
SFR larger than 1.8 \msunyr, the method allows to obtain the variation
of the bias $b$ with the stellar mass and SFR.  We obtain a strong
trend of decreasing $b$ with the stellar mass of the host galaxies
(Fig. \ref{FigBIASconstraints}).
This result is robust even if the models used for star-forming galaxies
do not include the scatter that would create interactions and starbursts that may
occur during the evolution of galaxies. In fact, these events may  be
responsible for the enhanced SFR and dilluted metallicity in LGRB hosts as discussed in section \ref{sec:discu}.
A similar trend is found with the SFR.
However, the uncertainties are larger in this case: contrary to
the stellar mass function, the SFR distribution evolves with redshift,
and suffers larger uncertainties.
We do not find any trend between $b$ and the SSFR (LGRBs are found in galaxies
with higher SSFR but $b$ does not depend on it), although our dynamical
range is too small to obtain a definitive answer.

The trends found between the bias $b$ and the stellar mass or the SFR are not
demonstrating a physical influence of these parameters on the occurrence
of LGRBs as these are ``unknown'' to massive stars that may explode as LGRBs. 
The properties of SFGs allow us to indirectly relate them to the metallicity
that on the other hand is likely to play an important role in the final 
stages of stellar evolution and thus on the occurrence of LGRBs. 
We obtain a clear 
trend of decreasing bias with increasing metallicity (Fig. \ref{FigBIASmetal}), 
that can put constraints
on the physics of LGRB progenitors.
Part of the relation that we obtained between the bias $b$ and the stellar mass,
the SFR, and the metallicity could also be due to the fact that dark bursts are missed in 
usual studies of LGRB host galaxies. 
A better determination of the properties of dark bursts hosts \emph{in
  the same redshift range} as our study could help to distinguish
between the intrinsic trend of $b$ with the metallicity or an observational 
bias (massive, metal-rich galaxies missing from our samples).  

In summary, despite large uncertainties and low statistics, this first application of our
method suggests a clear trend of decreasing bias $b$ with increasing 
metallicity, that may constrain the models of LGRB progenitors (role of metallicity 
in stellar evolution) or their environment (dust attenuation).
We are confident that this method will become much more powerful in the future, as 
i) the problem of dark bursts will be alleviated by robotic telescopes operating in the near IR
ii) larger samples of LGRB host galaxies detected below redshift unity will be
characterised, allowing us to use better controlled samples and  split the data into narrower redshift
intervals, limiting the evolutionary effects,
iii) it will also be possible to extend the method to higher redshifts with the determination of 
stellar mass functions and  properties  of SFGs in the early universe

\begin{acknowledgements}
This research has made use of the GHostS database (www.GRBhosts.org), which is partly 
funded by Spitzer/NASA grant RSA Agreement No. 1287913.
It was performed in the context, and with the support of the CNRS GDRE  ``Exploring the 
Dawn of the Universe with the GRBs'' (http://lamwws.oamp.fr/gdre/GdreGRBs).
We thank the anonymous referee for constructive comments.
S.B. thank the organisers of the ``Chemical evolution of GRB host galaxies'' 
Sesto workshop (F. Matteucci and F. Longo) and of the 5th GDRE 
workshop in IAP, as well as the participants to these events.
The suggestions made by the participants helped to improve this work. 
S.B. thanks everybody at the Trieste Observatory for welcoming him for 
several periods during which part of this work was performed, and finally
for discussion related to this work: O. Cucciati, S. Heinis, O. Ilbert.
\end{acknowledgements}

\end{document}